\documentclass[
    prb,
    twoside,
    twocolumn,
    10pt,
    floatfix,
    showpacs,
    citeautoscript,
]{revtex4-2}

\usepackage{amsmath}
\usepackage{amssymb}
\usepackage{color}
\usepackage{comment}
\usepackage{glossaries}
\usepackage{graphicx}
\usepackage[version=4]{mhchem}
\usepackage{siunitx}
\usepackage{times}
\usepackage{units}

\usepackage[
    unicode,
    colorlinks=true,
    urlcolor=blue,
    linkcolor=blue,
    citecolor=blue
]{hyperref}

\newacronym{aic}{AIC}{Akaike information criterion}
\newacronym{bic}{BIC}{Bayesian Information Criterion}
\newacronym{bz}{BZ}{Brillouin zone}
\newacronym{ce}{CE}{cluster expansion}
\newacronym{dft}{DFT}{density-functional theory}
\newacronym{eci}{ECI}{effective cluster interaction}
\newacronym{fcc}{FCC}{face-centered cubic}
\newacronym{fu}{f.u.}{formula unit}
\newacronym{mc}{MC}{Monte Carlo}
\newacronym{mcmc}{MCMC}{Markov chain Monte Carlo}
\newacronym{sgc}{SGC}{semi-grand canonical}
\newacronym{sro}{SRO}{short-range order}
\newacronym{vcsgc}{VCSGC}{variance-constrained semi-grand canonical}

\newcommand{\kk}{\boldsymbol{k}}
\newcommand{\caincoh}{c^\text{incoh}_\alpha}
\newcommand{\cbincoh}{c^\text{incoh}_\beta}
\newcommand{\cacoh}{c^\text{coh}_\alpha}
\newcommand{\cbcoh}{c^\text{coh}_\beta}

\newcommand{\etal}{\textit{et al.}}

\begin{document}
\title{
    Quantitative predictions of thermodynamic hysteresis:\texorpdfstring{\\}{}
    Temperature-dependent character of the phase transition in Pd--H
}

\author{J. Magnus Rahm}
\author{Joakim Löfgren}
\author{Paul Erhart}
\email{erhart@chalmers.se}
\affiliation{
  Chalmers University of Technology,
  Department of Physics,
  S-412 96 Gothenburg, Sweden
}

\begin{abstract}
The thermodynamics of phase transitions between phases that are size-mismatched but coherent differs from conventional stress-free thermodynamics.
Most notably, in open systems such phase transitions are always associated with hysteresis.
In spite of experimental evidence for the relevance of these effects in technologically important materials such as Pd hydride, a recipe for first-principles-based atomic-scale modeling of coherent, open systems has been lacking.
Here, we develop a methodology for quantifying phase boundaries, hysteresis, and coherent interface free energies using density-functional theory, alloy cluster expansions, and Monte Carlo simulations in a constrained ensemble.
We apply this approach to Pd--H and show that the phase transition changes character above approximately 400\,K, occurring with an at all times spatially homogeneous hydrogen concentration, i.e., without coexistence between the two phases.
Our results are consistent with experimental observations but reveal aspects of hydride formation in Pd nanoparticles that have not yet been accessible in experiment.
\end{abstract}

\maketitle

\section{Introduction}

\begin{figure*}
    \centering
    \includegraphics{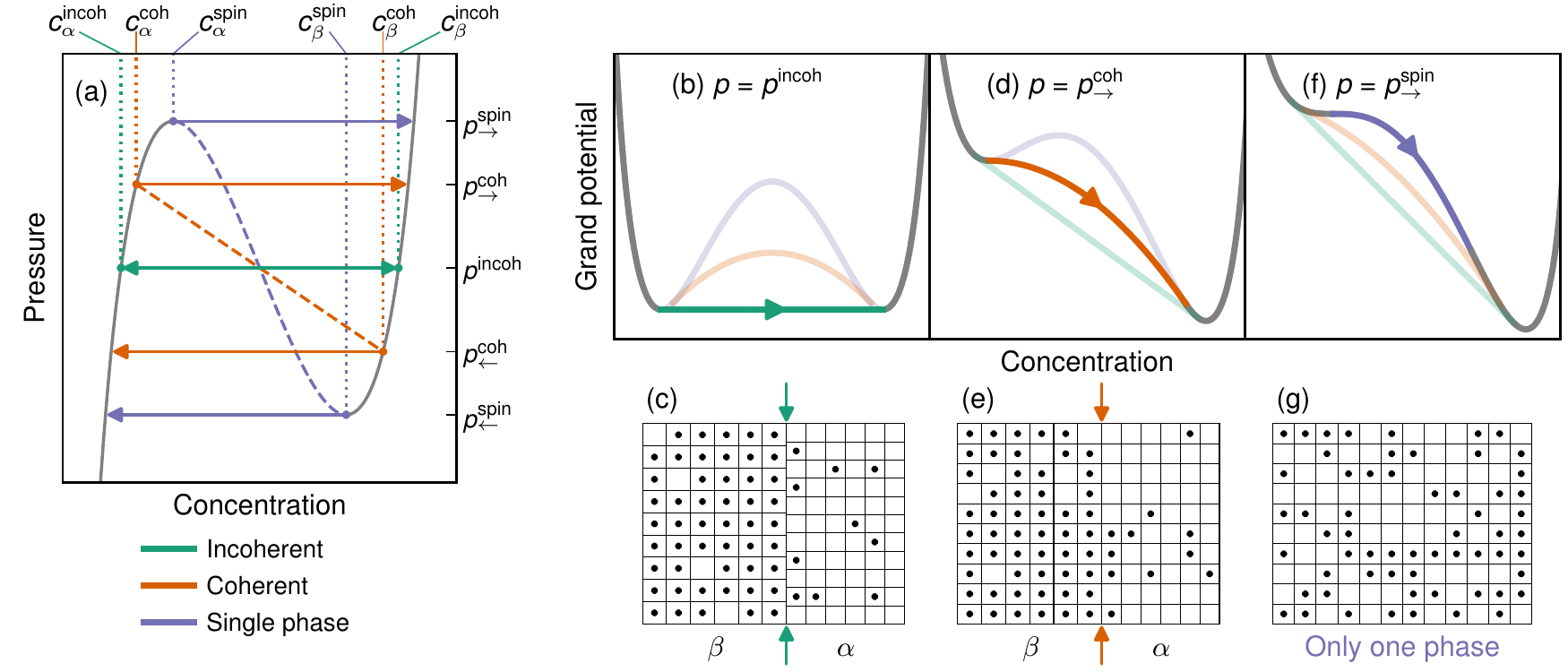}
    \caption{Pathways for a phase transition in a generic open system.
    (a) The phase transition is characterized by a plateau in the pressure--composition isotherm.
    The plateau occurs, however, at different pressures depending on the pathway of the phase transition.
    (b)~If the $\alpha$/$\beta$ interface is incoherent (as schematically depicted in (c)), the phase transition is thermodynamically allowed at $p=p^\text{incoh}$, where the two minima in the grand potential are equal.
    (d)~If the $\alpha$/$\beta$ interface remains coherent during absorption (as in (e)), a higher pressure, $p=p_\rightarrow^\text{coh}$, is required to overcome the strain energy barrier and induce the transition from $\alpha$ to $\beta$ phase.
    (f)~Absorption can in principle also occur with a continuously increasing but spatially homogeneous concentration (g), i.e., without the formation of two coexisting phases, if the pressure is increased to $p=p_\rightarrow^\text{spin}$ (the spinodal).
    Upon desorption, the incoherent phase transition occurs at the same pressure as absorption, whereas in the coherent and single-phase phase cases, lower pressures are required, leading to (thermodynamic) hysteresis.
    Note that interface energies and kinetic effects (especially related to nucleation), which are ignored here, will generally lead to some degree of (kinetic) hysteresis also in the incoherent case.
    }
    \label{fig:schematic-energy-landscape}
\end{figure*}

Phase transitions involving size-mismatched phases are ubiquitous in materials science.
Often such phase transitions are studied with the underlying assumption that the interface between the two phases is incoherent, such that the system is free from stress far from the interface.
There are, however, important cases for which this assumption breaks down.
Notable examples include intercalation in \ce{LiFePO4} nanoparticles used in Li-ion battery cathodes \cite{ZhaHulSin15} and hydride formation in Pd nanoparticles.
The latter has attracted increasing attention during the last decade, not only because Pd--H is a prototypical system for intercalation of small solute atoms in a host metal, but also due to its technological relevance in hydrogen storage \cite{SchWhiKan18}, optical hydrogen sensing \cite{WadSyrLan14, DarNugLan20}, and membrane reactors \cite{RahSamBab17}.
Mounting evidence \cite{GriStrGie16, UlvWelCha17, NarHayBal17} suggests that the phase transition from the hydrogen-poor $\alpha$ phase to the hydrogen-rich $\beta$ phase in Pd nanoparticles smaller than about \unit[300]{nm} in diameter occurs with a coherent interface between the two phases, in spite of their significant mismatch in lattice parameter.
As a consequence, the system exhibits considerable strain, which fundamentally alters the thermodynamics of phase transitions, as first shown in seminal works by Cahn, Larché, and others \cite{Cah62, LarCah73, LarCah78, Wil80, Wil84, CahLar84, JohVoo87, LiuAgr90}.
In particular, Schwarz and Khachaturyan \cite{SchKha95, SchKha06} have shown that in open systems, the strain energy associated with the coherent interface constitutes a macroscopic energy barrier that is inevitably associated with hysteresis, i.e., a hysteresis mandated by thermodynamics and insurmountable by thermal fluctuations (\autoref{fig:schematic-energy-landscape}).

The strive to better understand the dynamics of the phase transition in Pd--H nanoparticles has largely been driven by experiment \cite{WanFlaKuj02, BarHedPin13, WadPinOls14, BalNarKoh14, LanZhaZor10, SyrWadNug15}, notably including advanced imaging techniques that have revealed the details of the phase transition in time and space \cite{UlvWelCha17, NarHayBal17, SytHayNar18}.
Due to the challenges associated with transmission electron microscopy at high temperatures, atomic-scale imaging of nanoparticles during the phase transition has, however, generally been limited to temperatures well below room temperature, and thereby much lower than the temperature ranges where hydrogen sensors are required to operate.
As we will see below, this is unfortunate given the temperature-dependent character of the phase transition.
Meanwhile, modeling of coherent systems has primarily been based on phase-field models and other models with effective free energy expressions \cite{CogBaz12, CogBaz13, TanKar12, PhaGheCha19, PhaGheCha20, CheLiDu12, SpaGobHut16}, ground-state data assuming ideal mixing \cite{DoaWol12} or atomic-scale studies at specific temperatures \cite{SumTanIke15}.
At the same time, robust methodologies for first-principles calculations of thermodynamic properties on the atomic scale have remained insufficiently developed.
Thus, there have been few or no attempts to quantify important aspects such as the magnitude of the hysteresis and coherent interface energies from first principles, which may otherwise rationalize experimental observations in terms of the underlying atomic-scale interactions.

Here, we remedy this situation by demonstrating a methodology that allows us to calculate the phase diagram, the hysteresis, and the interface energies for open, coherent systems using a model built on first-principles calculations.
Notably, we obtain not only the incoherent but also the coherent and (incoherent) spinodal phase boundaries, and we demonstrate that the pathway of the phase transition changes character from coherent coexistence during the transition (\autoref{fig:schematic-energy-landscape}d--e) to a spatially homogeneous single-phase configuration above approximately \unit[400]{K} (\autoref{fig:schematic-energy-landscape}f--g).
Our approach is based on the use of \gls{dft} calculations for constructing an alloy \gls{ce} that explicitly takes strain into account, we sample the system using \gls{mc} simulations in a constrained ensemble that allows for free energy integration inside the two-phase region, and we demonstrate how to analyze these simulations in order to extract the relevant thermodynamic quantities.
While the individual parts of this approach are well-established, they have to the best of our knowledge not been combined previously.
We anticipate that the results will aid the interpretation and prediction of hydride formation in Pd nanoparticles.
The methodology is, however, not limited to Pd--H or even interstitial alloys, but can be applied to any alloy with a coherent phase transition between two phases with the same crystal structure.

\section{Theory}

\subsection{Phase separation in coherent systems} \label{sec:phase-separation-theory}

We begin with a general analysis of the energetics of coherently phase-separated systems to set the stage for the analysis of our simulations.
Consider a system with overall concentration $c_\text{av}$, phase separated coherently with interface orientation $\hat{\kk}$ into the $\alpha$ and the $\beta$ phases having, respectively, concentrations $c_\alpha$ and $c_\beta$.
The free energy of this phase-separated system can be written as
\begin{equation} \label{eq:free-energy-phase-separated}
\begin{split}
    F_{\alpha/\beta}(c_\text{av}, \hat{\kk})
    =
    \min_{c_\alpha, c_\beta} \big \lbrace
    \underbrace{V_\alpha f(c_\alpha) + V_\beta f(c_\beta)}_{\text{bulk free energy}} \\
    +
    \underbrace{V e_\text{strain}(c_\text{av}, c_\alpha, c_\beta, \hat{\kk})}_{\text{strain}}
    +
    \underbrace{A \gamma_{\alpha/\beta}(\hat{\kk}, c_\alpha, c_\beta)}_{\text{interface}}
    \big \rbrace.
\end{split}
\end{equation}
Here, $V_\alpha$ and $V_\beta$ are the volumes of the $\alpha$ and the $\beta$ phases (which can be determined from conservation of the number of atoms or atomic sites), $V$ the total volume ($V= V_\alpha + V_\beta$), $f(c)$ the free energy per volume of the single-phase system at concentration $c$, $e_\text{strain}(c_\text{av}, c_\alpha, c_\beta, \hat{\kk})$ the coherency strain energy per volume, $A$ the area of the interface, and $\gamma_{\alpha/\beta}(\hat{\kk}, c_\alpha, c_\beta)$ the interface energy.
If the strain and the interface energy terms are excluded, the expression is minimized by the well-known common tangent construction (\autoref{fig:schematic-energy-landscape}b, green line).
We will call the minimizing concentrations in this case the incoherent phase boundaries $c_\alpha^\text{incoh}$ and $c_\beta^\text{incoh}$, since they form the phase boundaries of the incoherently phase-separated system (in which the strain term vanishes).

In a coherently phase-separated system, the strain term does not vanish because the two phases need to adopt the same in-plane lattice parameter for the interface to be coherent.
While the interface energy term (scaling with area) becomes insignificant in a sufficiently large system, the strain term remains important since it scales with volume.
In general, the free energy $F_{\alpha/\beta}(c_\text{av}, \hat{\kk})$ is therefore \emph{not} minimized by the common tangent construction, but the concentrations $c_\alpha$ and $c_\beta$ may be, respectively, larger and smaller than $c_\alpha^\text{incoh}$ and $c_\beta^\text{incoh}$ \cite{SchKha95, SchKha06}.
We will refer to these minimizing concentrations as $c_\alpha^\text{coh}$ and $c_\beta^\text{coh}$ since they form the phase boundaries of the coherent system.
To summarize, in an incoherent system, there is phase separation for $c^\text{incoh}_\alpha < c_\text{av} < c^\text{incoh}_\beta$, whereas in a coherent system phase separation occurs for $c^\text{coh}_\alpha < c_\text{av} < c^\text{coh}_\beta$, and these concentrations satisfy the conditions
$c^\text{coh}_\alpha > c^\text{incoh}_\alpha$ and $c^\text{coh}_\beta < c^\text{incoh}_\beta$ (\autoref{fig:schematic-energy-landscape}a).

In open systems, the strain associated with coherent phase separation leads to a macroscopic energy barrier (\autoref{fig:schematic-energy-landscape}d), as opposed to the microscopic interface energy that constitutes the barrier in incoherent systems.
Such a macroscopic barrier cannot be overcome by fluctuations and therefore inevitably leads to a thermodynamic (i.e. non-kinetic) hysteresis \cite{SchKha95, SchKha06} (\autoref{fig:schematic-energy-landscape}a, orange lines).

Atomistic simulations are generally performed with simulation cells that are too small for interface contributions to be ignored.
To better understand this, we make the approximation that the interface energy is independent of the concentrations of the two phases over some range close to $c_\alpha$ and $c_\beta$, and write $V = AL$ (where $L$ is the length of the cell perpendicular to the interface).
We can compare the free energy of the phase-separated system, $F_{\alpha/\beta}(c_\text{av}, \hat{\kk})$, with the free energy of the single-phase system, $F_\text{sp}(c_\text{av})$.
Phase separation occurs if and only if $F_{\alpha/\beta}(c_\text{av}, \hat{\kk}) < F_\text{sp}(c_\text{av})$ and by rearrangement, using free energies per volume, this is equivalent to
\begin{equation} \label{eq:gamma-condition}
    \gamma_{\alpha/\beta}(\hat{\kk}) < L \left[ f_\text{sp}(c_\text{av}) - \tilde{f}_{\alpha/\beta}(c_\text{av}, \hat{\kk}) \right],
\end{equation}
where $\tilde{f}_{\alpha/\beta}(c_\text{av}, \hat{\kk})$ denotes the free energy per volume of the phase-separated system excluding the interface energy term.
If the strain energy term is sufficiently large, there will be no phase separation, because the expression on the right-hand side becomes negative.
We will show that this occurs above a certain critical temperature $T^\text{coh}_c$ in the Pd--H system.
We note, however, that even below this critical temperature, a cell sufficiently large in the direction perpendicular to the interface is needed for phase separation to be energetically favorable.
In a cell of finite length, phase separation will not occur in the full interval $c^\text{coh}_\alpha < c_\text{av} < c^\text{coh}_\beta$, but there will be intervals of concentrations above $c^\text{coh}_\alpha$ and below $c^\text{coh}_\beta$ where a single-phase system is formed.
The effect of interface energy in a small simulation cell is thus to narrow the two-phase region even further, although the concentrations of the constituent phases are unaffected (to a first approximation).
As shown below, the results from simulations using cells of finite size can, however, be extrapolated to yield the coherent phase boundaries in the thermodynamic limit.

\section{Methodology}
\label{sec:methods}

\subsection{DFT calculations} \label{sec:dft-calculations}
Energies were calculated with the projector augmented wave formalism as implemented in the Vienna ab initio simulation package (version 5.4.1, PAW 2015) \cite{KreFur96b, KreJou99} with the vdW-DF-cx exchange-correlation functional \cite{DioRydSch04, BerHyl14}.
Wave functions were expanded in a plane wave basis set with a cutoff of \SI{400}{\eV}, the \gls{bz} was sampled with a $\Gamma$-centered grid with a $\boldsymbol{k}$-point spacing of \SI{0.2}{\per\angstrom}, and occupations were set using Gaussian smearing with a width of \SI{0.1}{\eV}.
Atomic positions and cell shape were relaxed until residual forces were below \SI{25}{\milli\eV/\angstrom} and stresses below \SI{1}{\kilo\bar}.
Once converged, we ran an additional single-point calculation with a $\boldsymbol{k}$-point spacing of \unit[0.1]{\AA$^{-1}$} and the \gls{bz} integrations were carried out using the tetrahedron method with Bl\"{o}chl corrections.
In total, 368 different configurations were calculated, 30 of which were excluded from training and used as test set for the final \gls{ce} model.

\subsection{Alloy \texorpdfstring{\glspl{ce}}{CEs} with constituent strain}

In a regular alloy \gls{ce}, the energy of a structure with configuration $\mathbf{\sigma}$ is expanded in a sum of clusters $\alpha$, each associated with a an \gls{eci} $J_\alpha$,
\begin{equation} \label{eq:ce}
    E_\text{CE}(\mathbf{\sigma}) = J_0 + \sum_\mathbf{\alpha} J_\mathbf{\alpha} m_\mathbf{\alpha} \left \langle \Pi_{\alpha}(\mathbf{\sigma}) \right \rangle_{\alpha}.
\end{equation}
Here, $m_\alpha$ is the number of clusters that are symmetrically equivalent to $\alpha$ and $\left \langle \Pi_{\alpha}(\mathbf{\sigma}) \right \rangle_{\alpha}$ measures the average chemical order associated with the symmetrically equivalent clusters $\alpha$.
In our case, 
\begin{equation}
    \Pi_{\alpha} = \prod_{i\in\alpha} \Theta_i,
\end{equation}
where the product runs over all sites in the cluster and $\Theta_i = 1$ if site $i$ is occupied by a hydrogen atom, otherwise $-1$.
To construct a \gls{ce}, the \glspl{eci} $J_\alpha$ need to be fitted.
In practice, long-range clusters and clusters with many sites need to be truncated from Eq.~\eqref{eq:ce}.
Here, we included pairs no longer than 1.9 lattice parameters apart and triplets in which the interatomic distances were no longer than 1.75 lattice parameters.
The \glspl{eci} were obtained using automatic relevance determination regression \cite{scikit-learn} using the hyper parameter $\lambda = 10^4$.
To exactly reproduce pure Pd ($c=0$) and fully loaded Pd--H ($c=1$), we constrained the zerolet $J_0$ and the singlet $J_1$ during fitting.
The \gls{ce} was created using the \textsc{icet} package \cite{AngMunRah19}.

Long-ranged interactions mediated by strain are no longer captured after the \gls{ce} has been truncated.
In particular, the energetics of coherent interfaces in size-mismatched systems are qualitatively wrong.
To overcome this limitation of a standard alloy \gls{ce}, Laks \etal{} \cite{LakFerFro92} proposed the inclusion of an additional term referred to as the constituent strain,
\begin{equation} \label{eq:ce-and-strain}
\begin{split}
    E(\sigma) & = E_\text{CE}(\sigma) + E_\text{CS}(\sigma) \\
    & =E_\text{CE}(\sigma) + \sum_k \Delta E_\text{CS} (\hat{\kk}, c) F(\kk, \sigma).
\end{split}
\end{equation}
To find the \glspl{eci}, the second term is subtracted from the training data prior to fitting.
The constituent strain term associates an energy with static concentration waves, i.e., concentrations that vary along a direction $\hat{\kk}$.
Superlattices formed as a result of phase separation are examples of such concentration waves.
A superlattice is a structure in which $n$ atomic layers of phase A are followed by $n$ atomic layers of B, periodically repeated, and can be characterized by the orientation $\hat{\kk}$ of the interface between the two phases.
If the interface between the phases is coherent but the phases are size-mismatched, as is the case for the $\alpha$ and $\beta$ phases of Pd--H, such superlattices are associated with an elastic strain energy.
Since the stiffness is not isotropic, this energy depends on the orientation $\hat{\kk}$ of the coherent interface.
The term $\Delta E_\text{CS} (\hat{\kk}, c)$ is a material property that can be calculated from a large number of small \gls{dft} calculations using the same parameters as described above (but without relaxation), following the approach outlined by Ozolins \etal{} \cite{OzoWolZun98}.
It describes the excess energy (per atom) associated with superlattice formation in the $\hat{\kk}$ direction having an overall concentration $c$.
Specifically, it can be written
\begin{equation} \label{eq:ecs-def}
\begin{split}
    &\Delta E_\text{CS} (\hat{\kk}, c) = \\ &\quad\min_{a_\text{SL}} \left[ (1-c) \Delta E_\text{Pd}^\text{epi} (a_\text{SL}, \hat{\kk}) + c \Delta E_\text{PdH}^\text{epi} (a_\text{SL}, \hat{\kk}) \right],
\end{split}
\end{equation}
where the minimization is over the lattice parameter $a_\text{SL}$ in the plane of the interface and $E_X^\text{epi} (a_\text{SL}, \hat{\kk})$ represents the excess energy of phase $X$ when it is epitaxially strained to the lattice parameter $a_\text{SL}$.

The factor $F(\kk, \sigma)$ is a measure of the extent of configuration $\sigma$ matching a concentration wave with reciprocal vector $\kk$.
Following more recent versions of this approach \cite{LiuTriGia07, LiuZun08}, we write it as
\begin{equation} \label{eq:F}
    F(\kk, \sigma) = \left| S(\kk, \sigma) \right|^2 e^{- \left| \eta \kk \right|^2} \Big/ 4c(1-c),
\end{equation}
where $c$ is the concentration and $S(\kk, \sigma)$ is the structure factor.
The exponential term suppresses the constituent strain energy associated with rapidly varying concentrations, and we found that for the present system $\eta = 10$\,{\AA} provides an accurate \gls{ce} while the constituent strain is still well described.

Evaluation of the constituent strain involves a summation over $\kk$-points in the first Brillouin zone of the primitive \gls{fcc} cell, and nonzero $F(\kk, \sigma)$ are obtained for $\kk$ that are integer multiples of the reciprocal lattice vectors of the supercell.
For each nonzero value $F(\kk, \sigma)$, we need to know $\Delta E_\text{CS} (\hat{\kk}, c)$ to evaluate the strain energy.
For large simulation cells, this requires information about $\Delta E_\text{CS} (\hat{\kk}, c)$ for a large number of crystal directions $\hat{\kk}$.
To avoid an excessive number of \gls{dft} calculations, we used a fit to the known values of $\Delta E_\text{CS} (\hat{\kk}, c)$ to calculate constituent strain in unknown directions $\hat{\kk}$ (for details, see Supplementary Note~\ref{snote:fit-strain}).
Since low-index directions dominate the simulation cells employed in this work, the fitted values have only a marginal impact on the overall result. 

\subsection{Monte Carlo simulations in the VCSGC ensemble}
To extract thermodynamic quantities using our model, we performed \gls{mc} simulations using the \textsc{mchammer} module of the \textsc{icet} package \cite{AngMunRah19}.
To this end, we used a standard Metropolis algorithm in which the chemical identity of a randomly chosen site is flipped with probability
\begin{equation}
    P(\text{accept}) = \min \left \lbrace 1, \exp\left(- \Delta \psi / k_\text{B} T \right) \right \rbrace.
\end{equation}
The choice of $\psi$ corresponds to sampling of different thermodynamic ensembles.
In the present approach, we specifically need to access two-phase regions, where one value of the chemical potential maps to multiple concentrations.
We therefore used the \gls{vcsgc} ensemble \cite{SadErh12, SadErhStu12}, for which
\begin{equation}
    \psi = E + \bar{\kappa} N k_\text{B} T \left( c_\text{av} + \bar{\phi} / 2 \right)^2,
\end{equation}
where $E$ is the potential energy, $N$ the total number of sites, and $c_\text{av}$ the overall concentration.
The parameters $\bar{\phi}$ and $\bar{\kappa}$ constrain the average and the variance of the concentration, respectively.
\Gls{vcsgc} allows for thermodynamic integration, since the (canonical) free energy derivative is an observable of the ensemble,
\begin{equation}
    \frac{\partial F}{\partial c} = - 2 \bar{\kappa} N k_\text{B} T \left( \left \langle c_\text{av} \right \rangle + \bar{\phi} / 2 \right),
\end{equation}
where $\left \langle c_\text{av} \right \rangle$ is the observed overall concentration, in the present case the one of hydrogen.
This means that the free energy can be integrated even across phase boundaries.
Furthermore, the observed state can be related to the chemical potential difference of the constituents (here hydrogen and vacancies), since
\begin{equation}
    \bar{\mu}_\text{H} = \frac{1}{N} \frac{\partial F}{\partial c_\text{av}}.
\end{equation}
For the present case, this quantity is related to the absolute chemical potential of \ce{H2} gas according to $\mu_{\ce{H2}} = 2\left( \mu^0_\text{H} + \bar{\mu}_\text{H} \right)$, where $\mu^0_\text{H}$ is a reference that accounts for the offset introduced by fitting mixing energies rather than absolute energies.
 Assuming the \ce{H2} gas is ideal, we can relate the chemical potential to the partial pressure of \ce{H2} via
\begin{equation} \label{eq:ideal-gas-mu}
    \mu_{\ce{H2}}(T, p_{\ce{H2}}) = \mu^{\circ}_{\ce{H2}}(T) + k_\text{B} T \ln \frac{p_{\ce{H2}}}{p^{\circ}_{\ce{H2}}}.
\end{equation}
Since small errors in the temperature-dependent reference $\mu^{\circ}_{\text{H}_2}(T)$ have a large impact on the relation between $\mu_{\text{H}_2}$ and $p$, we chose to fit this reference level so as to reproduce the plateau pressure of the phase transition as measured in experiment (see \autoref{sfig:mu-to-pressure}).
This is the only experimental input in our model.
Our quantification of the hysteresis is, however, not impacted by this fit, only the absolute pressures for the transition from $\alpha$ to $\beta$ phase and vice versa.

\section{Results and discussion}

\subsection{Calculation of phase boundaries} \label{sec:single-phase-sim}
\begin{figure*}
    \centering
    \includegraphics{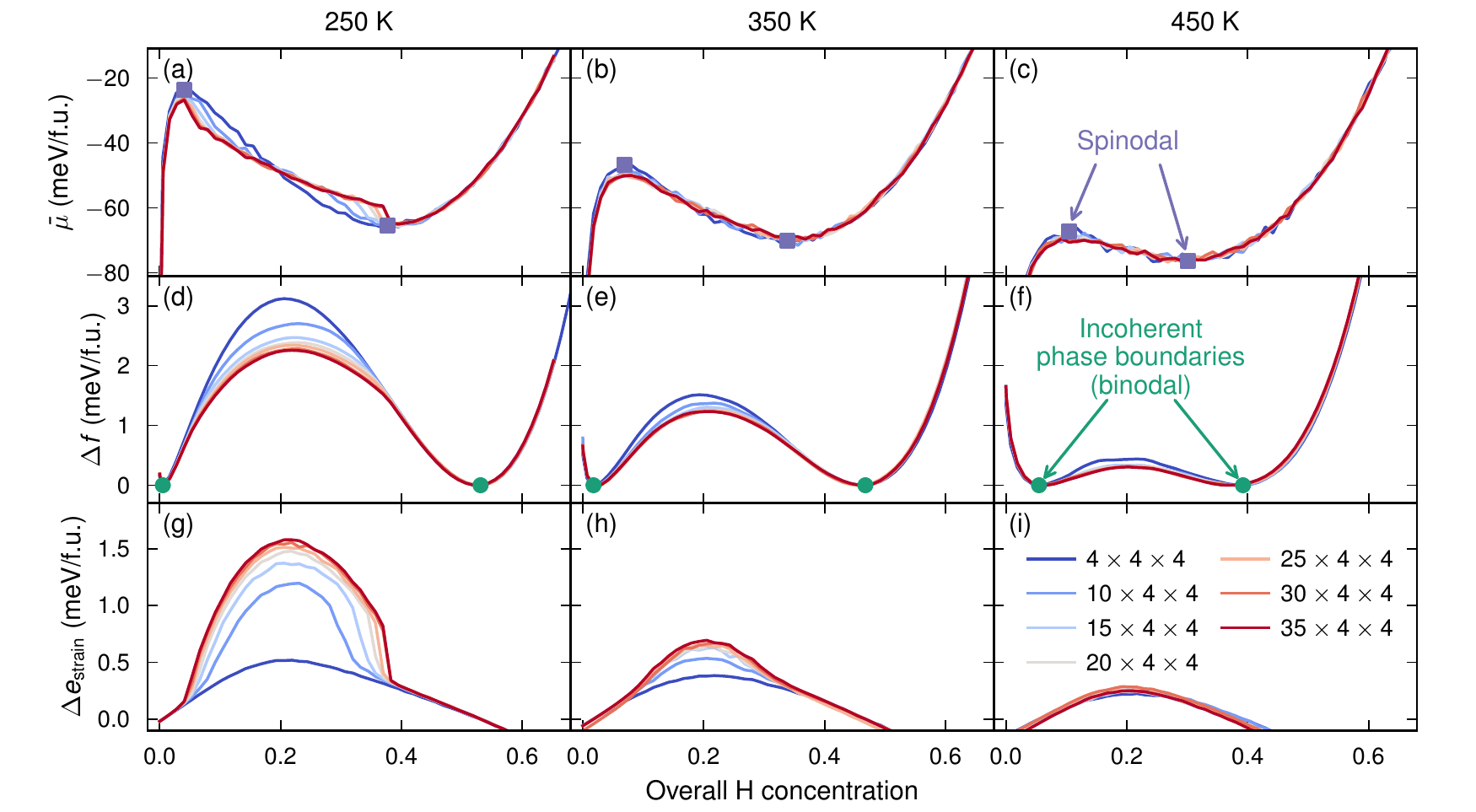}
    \caption{
    Energetics of Pd--H based on MC simulations at \unit[250]{K} (left), \unit[350]{K} (middle), and \unit[450]{K} (right) in cells elongated along $\langle100\rangle$.
    (a--c)~$\bar{\mu}$--$c$ isotherms always coincide above and below the phase boundaries, but in the two-phase region at low temperatures the isotherm changes with simulation cell length.
    (d--f)~The free energy decreases with increasing simulation cell length; yet it approaches a finite value inside the two-phase region, indicating a macroscopic energy barrier.
    The decrease with cell length is pronounced at low temperature but almost non-existent at high temperature.
    (g--i)~The free energy barrier results from a combination of chemical free energy and strain energy, but the strain energy contribution is significant only in long cells at low temperatures, indicating that it is only in these cells that phase separation occurs.
    Here, the free energy (d--f) and the strain energy (g--i) have been tilted so as to be zero at the incoherent phase boundaries.
    }
    \label{fig:100-thermo}
\end{figure*}

To calculate phase boundaries, we begin by studying the free energy landscape obtained from \gls{mc} simulations in simulation cells of increasing length, ranging in size from $4 \times 4 \times 4$ conventional (4-atom) unit cells (256 sites) to $4 \times 4 \times 35$ conventional unit cells (2,240 sites) using a \gls{ce} with strain (for details concerning the \gls{ce} model, see Supplementary Note~\ref{snote:performance-of-the-ce}, \autoref{sfig:ce-quality} and \autoref{sfig:superlattice-quality}).
Our approach allows us to obtain the relationship between hydrogen chemical potential $\bar{\mu}$ and overall concentration $c_\text{av}$ also in the two-phase $\alpha + \beta$ region, which is characterized by one value of $\bar{\mu}$ mapping to several values of $c_\text{av}$ (\autoref{fig:100-thermo}a--c).
By integration, we obtain the free energy $\Delta f(c_\text{av})$ (\autoref{fig:100-thermo}d--f, here tilted to make $\Delta f(c^\text{incoh}_\alpha) = \Delta f(c^\text{incoh}_\beta) = 0$).
The $\alpha + \beta$ two-phase region is manifested by a concave interval in $\Delta f(c_\text{av})$.
By drawing the convex hull, we can identify the \emph{incoherent} phase boundaries, $c^\text{incoh}_\alpha$ and $c^\text{incoh}_\beta$ (green circles in \autoref{fig:100-thermo}d--f).
These are the equilibrium concentrations of an incoherently phase-separated system.

The identification of incoherent phase boundaries (the binodal) is insensitive to simulation cell size and shape, since free energies calculated with different simulation cells only differ inside the two-phase region.
Nevertheless, these differences reflect important aspects of the simulations.
At low temperatures, the chemical potential in the two-phase region (\autoref{fig:100-thermo}a, \unit[250]{K}) is such that the free energy barrier becomes lower as the simulation cell grows longer (\autoref{fig:100-thermo}d).
This is expected, since with a larger cell the system has greater configurational freedom and can thus explore more configurations.
The main difference between the small, cubic cell and the long cells, is that the latter support phase separation.
This is most clearly manifested by a larger strain energy inside the two-phase region in long cells than in the small, cubic cell (\autoref{fig:100-thermo}g).
At \unit[350]{K} (\autoref{fig:100-thermo}h), these effects are much less pronounced and at \unit[450]{K} (\autoref{fig:100-thermo}i), they are non-existent; the strain energy is almost identical in the smallest and the largest cell.
This indicates that phase separation does not occur, which is readily confirmed by inspection of the trajectories.

The \emph{spinodal} lies inside the incoherent phase boundaries where the free energy depends on simulation cell size, and thus it cannot be deduced from an arbitrary simulation cell.
To calculate the spinodal, we track, respectively, the position of the local maximum and minimum in the $\bar{\mu}$--$c_\text{av}$ isotherm for the smallest simulation cell (purple squares in \autoref{fig:100-thermo}a--c), because this is the only cell that maintains the single-phase, homogeneous concentration profile required for the spinodal to be well-defined.

A few additional observations are helpful at this point.
The $\bar{\mu}$--$c_\text{av}$ isotherm (\autoref{fig:100-thermo}a) converges fairly quickly as the simulation cell becomes longer.
It does, however, not converge to the horizontal line connecting $\caincoh$ and $\cbincoh$ (the common tangent construction) but to a sloping line with a weak curvature.
Differences between cells of different length are largest close to the edges of the two-phase region, because phase separation is not necessarily favorable if the volume fraction of the minority phase is too small, an effect that becomes more pronounced the smaller the cell.
The system then adopts a single-phase configuration, and the thermodynamic equilibrium state coincides with the small $4\times4\times4$ cell.
As a result of the slope of the free energy derivative, the free energy barrier approaches a finite value; in other words, it does not vanish in the thermodynamic ($N\rightarrow\infty$) limit.
This finite energy barrier is a consequence of the system being \emph{coherently} phase separated.
These observations are consistent with the model by Schwarz and Khachaturyan \cite{SchKha95, SchKha06}.

\begin{figure}
    \centering
    \includegraphics{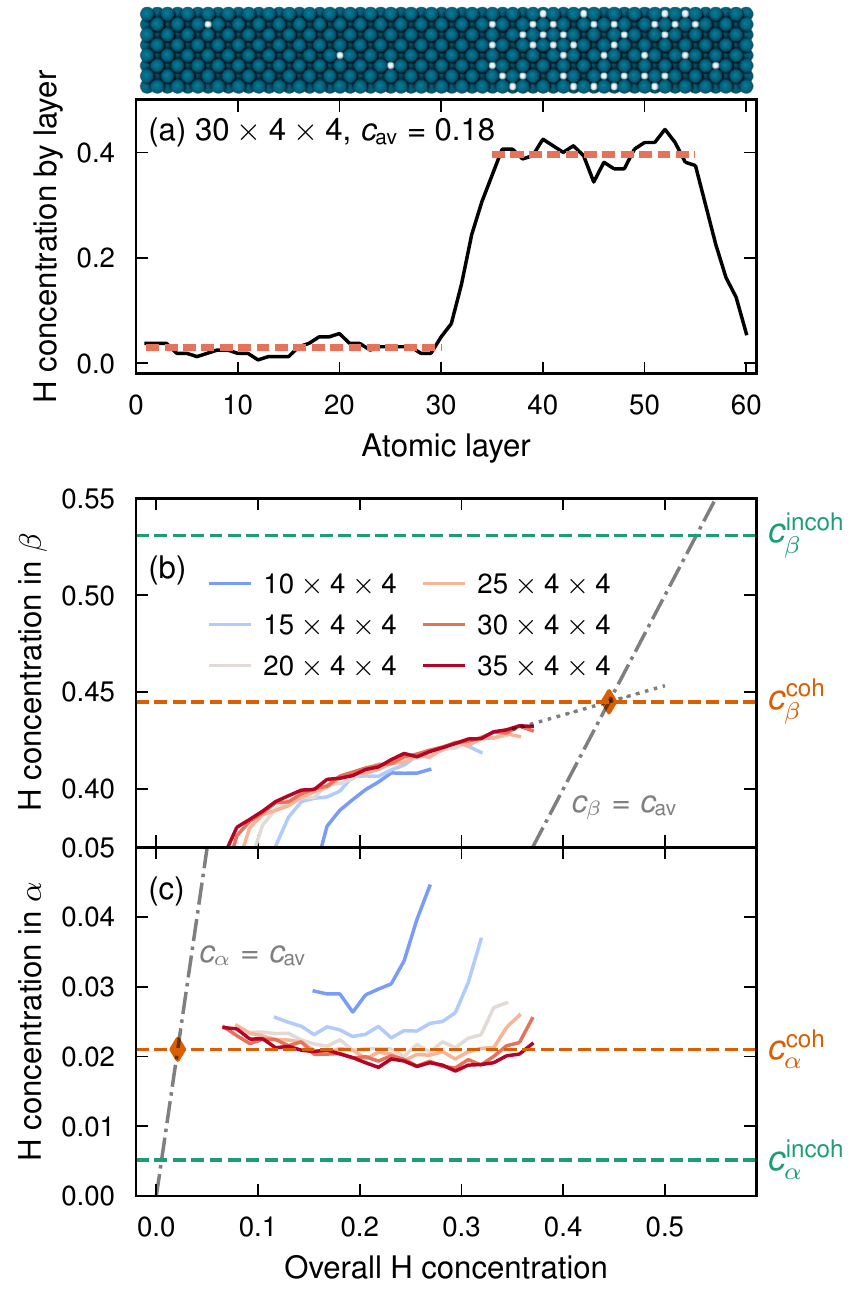}
    \caption{
    Extraction of coherent phase boundaries.
    Data presented concerns \{100\} interfaces at \unit[250]{K}, but the procedure is analogous for other interfaces and temperatures.
    (a)~The hydrogen concentration in each atomic layer was calculated and averaged over five neighboring layers (black line), in this case for a snapshot of a $30 \times 4 \times 4$ simulation cell, having 60 atomic layers (inset, note that Pd atoms are fixed).
    The concentration in the $\alpha$ and the $\beta$ phase was taken as the average concentration in the longest set of contiguous atomic layers that had, respectively, concentration below and above $0.2$ (red, dashed line).
    (b)~By averaging the thus obtained concentrations over all snapshots, using different overall concentrations, one observes that the concentration in the $\beta$ phase is independent of simulation cell size for sufficiently long cells if the overall concentration is not too close to the phase boundaries.
    The concentration further increases as the $\beta$ phase grows.
    By linear extrapolation (gray, dotted line) to 100\% volume fraction $\beta$, where the overall concentration is equal to the concentration in the $\beta$ phase, the coherent phase boundary $\cbcoh$ is extracted  (orange, dashed line).
    This concentration is significantly smaller than $\cbincoh$ (green, dashed line, as extracted in \autoref{fig:100-thermo}d--f).
    (c)~In the $\alpha$ phase, the concentration is essentially unaffected by the overall concentration.
    The corresponding coherent phase boundary $\cacoh$ (orange, dashed line) is significantly larger than the incoherent phase boundary $\caincoh$ (green, dashed line).
    }
    \label{fig:coherent-phase-boundaries}
\end{figure}

To extract the \emph{coherent} phase boundaries, we are interested in the compositions of the two phases in the coherently phase-separated system.
To extract these, we calculated a running average of hydrogen concentration per atomic layer, with the average taken over five layers.
We then calculated the hydrogen concentration in the $\alpha$ phase as the average concentration of the longest series of contiguous atomic layers that had a concentration below 20\%, and similarly for the $\beta$ phase (illustrated in \autoref{fig:coherent-phase-boundaries}a).
The results show that the hydrogen concentration of the $\beta$ phase increases as the $\beta$ phase grows and the overall concentration increases (\autoref{fig:coherent-phase-boundaries}b).
It is, however, always significantly smaller than $\cbincoh$.
In the $\alpha$ phase, the concentration is essentially constant but larger than $\caincoh$ (\autoref{fig:coherent-phase-boundaries}c).
As discussed above, phase separation is not necessarily favorable in finite simulation cells with overall concentrations close to the phase boundaries.
Consequently, our data for the composition of the $\alpha$ and the $\beta$ phases do not reach the points where the concentration of the phase equals the overall concentration (i.e., where the volume fraction of the phase is 100\%).
Nevertheless, the available data enables extrapolation to these points (orange diamonds in \autoref{fig:coherent-phase-boundaries}b--c).
These points yield the coherent phase boundaries at the corresponding temperature, $\cacoh$ and $\cbcoh$, because they mark the limits of the concentration interval in which coherent phase coexistence is favorable (compared to a single-phase configuration) in the thermodynamic limit.
We emphasize that the coherent and incoherent phase boundaries differ significantly.
We also note that, in the present system, it is sufficient to consider simulation cells with \{100\} interfaces when calculating coherent phase boundaries, since any other orientation leads to larger strain energy (see \autoref{sfig:superlattice-quality}) and thereby thermodynamically less favorable coherent phase coexistence.

\subsection{The Pd--H phase diagram}
\begin{figure}
    \centering
    \includegraphics{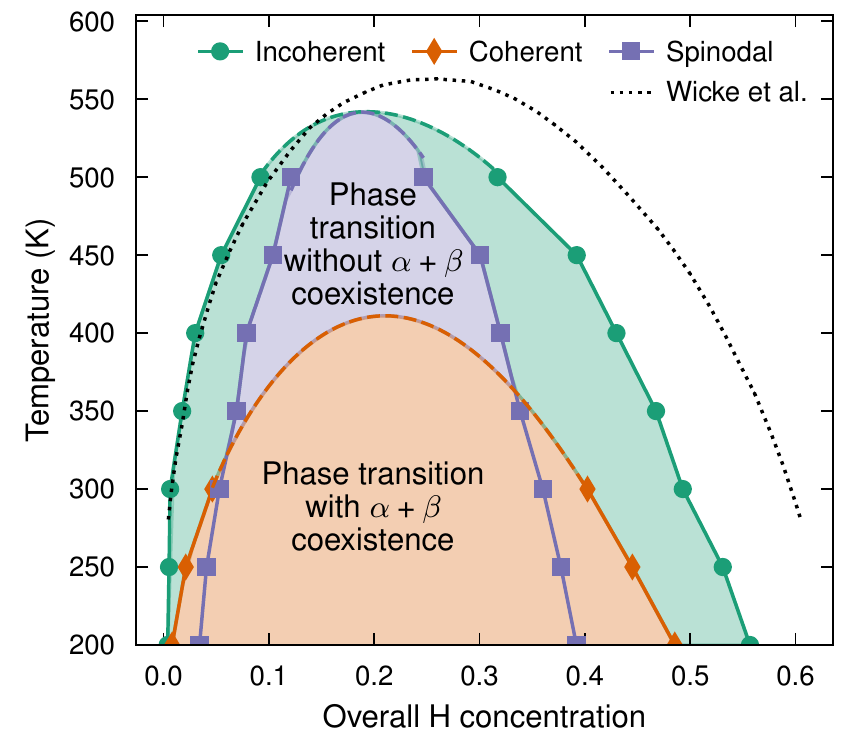}
    \caption{
    Phase diagram for Pd--H.
    The incoherent phase boundaries (binodals, green lines) indicate where the phase transition is thermodynamically favorable in the stress-free case.
    The critical temperature agrees well with experimental results (\unit[540]{K} in our model, \unit[560]{K} experimentally \cite{WicBla87}), and so does the phase boundary on the $\alpha$ side, while the concentration in the $\beta$ phase is underestimated compared to experiment.
    If the phase transition occurs fully coherently, phase coexistence between $\alpha$ and $\beta$ during the transition is only favorable between the coherent phase boundaries (orange area).  
    Above approximately \unit[400]{K}, coherent phase coexistence is no longer favorable for any composition, and the phase transition from $\alpha$ to $\beta$ and vice versa proceeds with a spatially homogeneous hydrogen concentration in the full concentration range.
    This phase transition becomes spontaneous within the incoherent spinodal (purple area).
    Dashed lines indicate extrapolation of the results of the simulations.
    }
    \label{fig:phase-diagram}
\end{figure}

By tracking the incoherent and coherent phase boundaries as well as the spinodal as a function of temperature, we can construct the phase diagram (\autoref{fig:phase-diagram}).
We note first that the critical temperature predicted by our model (\unit[540]{K}) agrees well with the experimental data by Wicke \etal{} \cite{WicBla87}, while noting that the critical temperature obtained with a \gls{ce} is in general associated with a significant statistical uncertainty \cite{AngMunRah19, RahLofFra21}.
The incoherent phase boundary (green line) marks the concentrations where the phase transition may start if it is fully incoherent.
Compared to experiment (taken from desorption isotherms of bulk samples, which are believed to be close to the actual incoherent phase boundaries \cite{WicBla87}), the phase boundary on the $\alpha$ side is in very good agreement, while we underestimate the concentration on the $\beta$ side.
These differences are likely related to both the (in)accuracy of the underlying \gls{dft} calculations and the treatment of temperature-dependent properties, notably the neglect of vibrations.
Similar results for the incoherent phase diagram have previously been obtained in \gls{ce} studies \cite{BouCenCri19, RahLofFra21}.
Interestingly these studies, which did not include a special treatment of strain, yielded a more pronounced underestimation of the critical temperature.

The coherent phase boundaries (orange line in \autoref{fig:phase-diagram}) mark the concentrations where fully coherent phase separation may start.
These are always inside the incoherent phase boundaries because of the additional strain energy contribution.
Coherent phase separation between $\alpha$ and $\beta$ becomes energetically unfavorable above approximately \unit[400]{K}, because then the strain energy associated with phase separation is always larger than the gain in chemical free energy.
We will refer to this as the coherent critical temperature.
The exact value of the coherent critical temperature is difficult to determine accurately and was here obtained by extrapolation of the coherent phase boundaries in the range of \unit[200--300]{K} where the constitution of the phases is still relatively easy to determine.
The value of \unit[400]{K} obtained by this procedure is further supported by an analysis of the histogram of atomic layer concentrations (\autoref{sfig:phase-constitution}).
Our calculated coherent critical temperature is higher than the only previous attempt to determine the coherent critical temperature in Pd--H that we are aware of \cite{HoGolWea79}.
Based on experimental data for elastic constants, lattice parameters, and incoherent critical temperature, the latter gave an estimate for the coherent critical temperature between approximately 260 and \unit[350]{K}.

We thus predict that there is an interval between approximately \unit[400]{K} and \unit[540]{K} where the phase transition occurs without $\alpha + \beta$ coexistence.
In this interval, the phase transition does not proceed by nucleation and growth, since nucleation would be associated with a strain energy that exceeds the gain in chemical free energy.
Instead, the system will always remain in a single-phase configuration (as schematically indicated in \autoref{fig:schematic-energy-landscape}g).
We emphasize that the the transition from $\alpha$ to $\beta$ (and vice versa) is still a first-order phase transition in this temperature interval, and is as such characterized by a plateau in the absorption/desorption isotherm as well as hysteresis.
The behavior of the system above and below \unit[400]{K} only differs with regard to the \emph{pathway} of the phase transition. 

We have also drawn the incoherent spinodal in \autoref{fig:phase-diagram} (purple line).
The significance of the incoherent spinodal in an open, coherent system is that it marks the concentrations where the chemical potential of hydrogen is large (small) enough for any free energy barrier for absorption (desorption) to be eliminated (as indicated in \autoref{fig:schematic-energy-landscape}f).
In other words, the path from $\alpha$ to $\beta$ or vice versa becomes monotonically downhill in the grand potential at the chemical potential corresponding to these concentrations.
Hence, the phase transition will always occur spontaneously once the incoherent spinodal has been reached.
(We also acknowledge the existence of the coherent spinodal, where the coherent system becomes unstable with respect to formation of concentration waves, but it is not relevant for our analysis.)

\subsection{Calculation of hysteresis}
\begin{figure}
    \centering
    \includegraphics{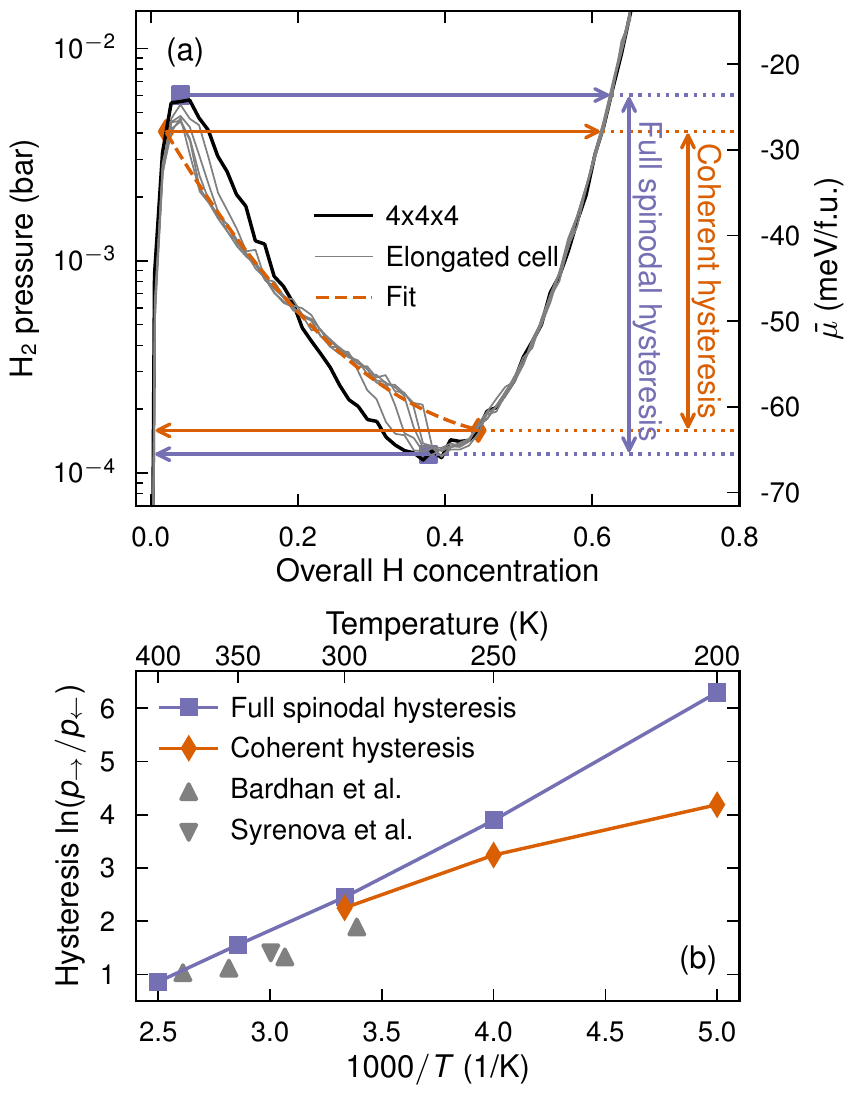}
    \caption{
    Thermodynamic hysteresis in the coherent Pd--H system.
    (a)~$p$--$c$ isotherm at \unit[250]{K}.
    We identify two kinds of hysteresis, full spinodal hysteresis (purple lines) and coherent hysteresis (orange lines), i.e., the thermodynamic hysteresis intrinsic to the coherent phase transition.
    The former is derived by finding the local maximum and minimum in the isotherm, whereas the latter requires extrapolation to the thermodynamic (infinite size) limit (orange, dashed line).
    (b)~Coherent hysteresis is smaller than full spinodal hysteresis as long as the coherent phase boundary lies outside the spinodal ($\lesssim \unit[350]{K}$), but for higher temperatures the hysteresis will always be full spinodal (unless dislocations form).
    Compared to measurements on nanoparticles \cite{BarHedPin13, SyrWadNug15} (gray triangles), we make a slight overestimation of the hysteresis, possibly because of defect formation during dehydration in experiment.
    Hysteresis is here quantified as the logarithm of the ratio between the transition pressure from $\alpha$ to $\beta$ ($p_\rightarrow$) and $\beta$ to $\alpha$ ($p_\leftarrow$).
    }
    \label{fig:hysteresis}
\end{figure}

Equipped with the coherent phase boundaries, we can now extrapolate the coherent energy landscape to the thermodynamic limit.
To this end, we fit a second degree polynomial (orange, dashed line in \autoref{fig:hysteresis}a) to $\bar{\mu}$ as a function of $c_\text{av}$ for the longest cells in the concentration interval [0.1, 0.3], while enforcing that the resulting curve crosses the single-phase $\bar{\mu}$--$c_\text{av}$ curve (black line in \autoref{fig:hysteresis}a) at the concentrations $\cacoh$ and $\cbcoh$ calculated in the previous section.
The (thermodynamic) hysteresis is then given by the pressure difference between the highest and lowest points on this line; this is the path of lowest free energy in the thermodynamic limit.
Below approximately \unit[350]{K}, this hysteresis is smaller than the full spinodal hysteresis, which is given by the difference between the maximum and the minimum of the black line in \autoref{fig:hysteresis}a.
When the spinodal lies outside the coherent phase boundaries, the system will always undergo full spinodal hysteresis; the system cannot lower its free energy by phase separation outside the coherent phase boundaries.
Using the full spinodal hysteresis as an approximation of the coherent hysteresis, as has previously been done in the literature \cite{GriStrGie16, SchKhaCar20}, is thus sensible if the temperature is sufficiently high.

We quantify the hysteresis using $\ln p_\rightarrow / p_\leftarrow$, where $p_\rightarrow$ and $p_\leftarrow$ are the transition pressures from $\alpha$ to $\beta$ phase and vice versa (\autoref{fig:hysteresis}b).
We note that our hysteresis is significantly narrower than the hysteresis predicted by Schwarz \etal{} \cite{SchKhaCar20} (for example, at \unit[300]{K} we obtain a coherent hysteresis $\ln p_\rightarrow / p_\leftarrow \approx 2.3$, which is less than half of what was predicted in \cite{SchKhaCar20}).
We attribute the majority of this difference to the choice of the underlying atomic scale interaction model, as a semi-empirical embedded atom method potential was used in \cite{SchKhaCar20}.

Comparison of our results to the hysteresis experimentally measured in Pd nanoparticles smaller than \unit[300]{nm}, which are widely considered to transition coherently, is not entirely straight-forward, since the hysteresis decreases with the size of the nanoparticles.
This is connected to the critical temperature also decreasing with smaller nanoparticle size \cite{GriStrGie16}, an effect that our model does not take into account.
To enable a fair comparison, we extracted experimental data for two cases where the hysteresis was measured at fixed temperature for single-crystalline nanocubes of different sizes \cite{BarHedPin13, SyrWadNug15}, and extrapolated the size-dependent hysteresis to the bulk limit (gray triangles in \autoref{fig:hysteresis}b, extrapolation in \autoref{sfig:hysteresis-experimental-extrapolation}).
Our model still slightly overestimates the hysteresis compared to these data.
Previous analyses have indicated that the $\alpha \rightarrow \beta$ phase transition occurs fully coherently in Pd nanoparticles, whereas this does not always apply for the $\beta \rightarrow \alpha$ transition \cite{GriStrGie16, NarHayBal17}.
If desorption is at least partially incoherent, the hysteresis will shrink, which may explain the discrepancy between the hysteresis in our model and experiment, as our model assumes fully coherent phase transitions during both loading \emph{and} unloading.
It should further be noted that the surfaces of nanoparticles allow for stress relaxation, in particular far from the $\alpha/\beta$ interface, which most likely leads to a lowering of the free energy barrier of the phase transition associated with coherency stresses.
It is possible that the extrapolation to the bulk limit corrects for this effect to some extent, but not necessarily completely.
The overestimation of the hysteresis is very slight and is thus not necessarily a sign that dislocations form during hydration.

Finally, we note that the results in \autoref{fig:hysteresis} were obtained with simulations cells with \{100\} interfaces; any other orientation would yield a larger coherent hysteresis, since the strain is the smallest for interfaces with \{100\} interfaces (as quantified in \autoref{sfig:superlattice-quality}).
This does not necessarily mean that the phase transition will always proceed with a \{100\} phase front, because interface energies may be smaller in other directions.
Interface energies are, however, by definition not taken into account for the thermodynamically mandated coherent hysteresis.

\subsection{Calculation of interface free energies}
\begin{figure}
    \centering
    \includegraphics{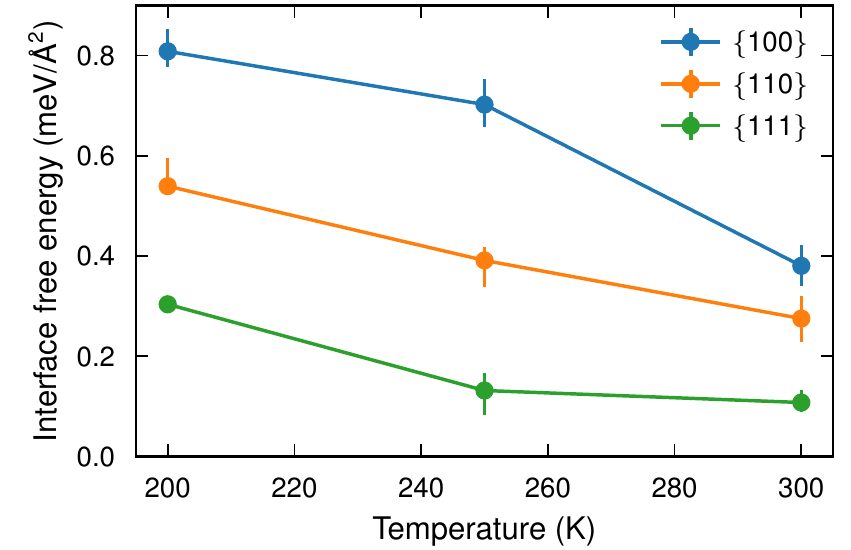}
    \caption{
    Coherent $\alpha$/$\beta$ interface free energies in Pd--H.
    The interface energies are the largest for interfaces with \{100\} orientation and smallest for \{111\}.
    The error bars represent the spread of interface energies over the (overall) concentration interval [0.15, 0.25], and the symbols the average in that interval.
    }
    \label{fig:interface-energies}
\end{figure}

While interface free energies do not impact coherent hysteresis in the thermodynamic limit, they are in practice expected to play a role for the actual phase transition, as they constitute a microscopic energy barrier that needs to be overcome.
Our simulations allow us to extract the coherent interface free energies as well (for methodological details, see Supplementary Note~\ref{snote:calculation-of-interface-energies} and \autoref{sfig:interface-energies-extrapolation}).
The results show that the interface energy of \{100\} is larger than both \{111\} and \{110\} (\autoref{fig:interface-energies}).
Hence, while strain favors a phase transition dominated by \{100\} interfaces, the (chemical) interface energies favor other interface orientations.
It should be noted that the lower interface energies of \{111\} and \{110\} are not necessarily only an effect of favorable chemistry.
The concentrations of the constituent phases vary between the three interface orientations, since different interfaces strike different balances between chemistry and strain energy.
Nevertheless, the impact of interface energies is to stabilize \{111\} and \{110\} interfaces relative to \{100\}, which may provide a clue as to why the former interfaces have been routinely observed experimentally during hydrogenation of Pd nanoparticles \cite{SytHayNar18}.
We note, however, that the interface energies are small in absolute terms; unless the nucleus size is less than a few nanometers, the free energy barrier (\autoref{fig:100-thermo}d) will be completely dominated by strain.

\section{Conclusion}
We have found that the combination of alloy \glspl{ce} including strain and \gls{mc} simulations in a constrained ensemble with cells of different sizes and shapes, constitutes a robust methodology for extracting the most relevant thermodynamic quantities of coherent phase transitions in an open system.
Most importantly the present approach enables one to quantitatively predict not only the incoherent phase boundaries and the (incoherent) spinodal, but also the coherent phase boundaries, using only input from first-principles calculations.
The results are consistent with the theory of Schwarz and Khachaturyan \cite{SchKha95, SchKha06}, with small qualitative differences that can be explained by small deviations from the assumptions in this theory.
Specifically, these assumptions implied that the compositions of the two phases in the coherently phase-separated system were independent of the overall concentration.
The variation of the composition of the $\beta$ phase with overall concentration that we observe (\autoref{fig:coherent-phase-boundaries}b) is likely the result of a small but significant deviation from Vegard's law (\autoref{sfig:vegard}) as well as the concentration dependence of the elastic constants (\autoref{sfig:elastic-constants}), which have been shown to impact the dependence of phase composition on the overall composition \cite{PfeVoo91, LeeTao94, CheLiDu12, KorPezBri18}. 

The phase diagram for Pd--H computed here (\autoref{fig:phase-diagram}) reveals that there is an interval between approximately \unit[400]{K} and \unit[540]{K} where the phase transition proceeds with a hydrogen distribution that is at all times homogeneous throughout the material, without any formation of two coexisting phases.
To the best of our knowledge, no direct, experimental observations of such a phase transition in Pd--H has been reported yet, which is not surprising given that \textit{in situ} imaging has been restricted to much lower temperatures \cite{SytHayNar18, NarHayBal17}.
It does, however, urge caution: conclusions drawn about the dynamics of the phase transition at low temperatures may not be applicable at higher temperatures, since they occur by different processes.

Our approach further allows for quantification of the hysteresis, both above \unit[350]{K} where it is fully spinodal, and below \unit[350]{K} where it is anywhere between fully spinodal and coherent.
The results are largely consistent with available experimental data, although comparison is not straight-forward.
We reemphasize that the present results are applicable to fully coherent phase separation in Pd--H and thus mostly applicable to nanoparticles with diameters below \unit[300]{nm} \cite{GriStrGie16, UlvWelCha17}.
We note that if defects (notably dislocations) are formed, which is expected in bulk Pd--H, other models such as the one in Ref.~\cite{WeaVooFul21} are better suited to describe hysteresis \cite{WeaVooFul21}.
The latter model, however, relies on interface pinning and is therefore not suitable to analyze hysteresis in single-crystalline nanoparticles. 

Finally, using extrapolation of data from simulation cells with varying shapes, interface free energies can be calculated.
In the present case, the interface free energies are very small, which is not unexpected given that a condition for coherent phase transitions to be competitive, is that their interface free energies are lower than the incoherent interface free energies.
For most purposes, the interface free energies of Pd--H are small enough to be safely ignored; the direction-dependent strain energy is by far the dominant term unless the nucleus size is smaller than a few nanometers.
Yet, these interface energies may play a role in the pathway of the hydration of Pd nanoparticles, which has been observed to occur, at least sometimes, via interface geometries disfavored by strain but favored by interface free energies \cite{SytHayNar18}.

\section*{Acknowledgements}
This work was funded by the Knut and Alice Wallen\-berg Foundation (grant number 2015.0055), the Swedish Research Council (grant numbers 2015-04153, 2018-06482, 2020-04935), and the Swedish Foundation for Strategic Research (grant number RMA15-0052).
The computations were enabled by resources provided by the Swedish National Infrastructure for Computing (SNIC) at NSC, HPC2N and PDC partially funded by the Swedish Research Council (grant number 2018-05973).
We thank Prof. Mark Asta and Dr. Babak Sadigh for insightful discussions.

\end{document}